\documentclass[aps,eqsecnum,amsmath,twocolumn, preprintnumbers]{revtex4}
\usepackage{graphics,graphicx,setspace,epsfig,color}
\usepackage[vcentermath]{}
\usepackage{epsf}
\usepackage{amscd}
\usepackage{amsmath,amssymb}

\newcommand{\dslash}{D\!\!\!\!\slash}
\newcommand{\beq}{\begin{equation}}
\newcommand{\eeq}{\end{equation}}
\newcommand{\bea}{\begin{eqnarray}}
\newcommand{\eea}{\end{eqnarray}}
\newcommand{\nn}{\nonumber}

\newcommand{\ded}{{\mathbf{d}^\dagger}}
\newcommand{\de}{{\mathbf{d}}}
\newcommand{\ab}{\mathbf{a}}
\newcommand{\bb}{\mathbf{b}}

\begin{document}

\preprint{UMD-40762-482}

\title[title]{The phases of deuterium at extreme densities}
\author{Paulo F. Bedaque}
\email{bedaque@physics.umd.edu}
\affiliation{Maryland Center for Fundamental Physics \\ 
Department of Physics\\
University of Maryland\\
College Park, MD 20742}

\author{Michael I. Buchoff}
\email{mbuchoff@physics.umd.edu}
\affiliation{Maryland Center for Fundamental Physics \\ 
Department of Physics\\
University of Maryland\\
College Park, MD 20742}

\author{Aleksey Cherman}
\email{alekseyc@physics.umd.edu}
\affiliation{Maryland Center for Fundamental Physics \\ 
Department of Physics\\
University of Maryland\\
College Park, MD 20742}


\begin{abstract}
We consider deuterium compressed to higher than atomic, but lower than nuclear densities.  At such densities deuterium is a superconducting quantum liquid.  Generically, two superconducting phases compete, a ``ferromagnetic" and a ``nematic" one. We provide a power counting argument suggesting that the dominant interactions in the deuteron liquid are perturbative (but screened) Coulomb interactions.  At very high densities the ground state is determined by very small nuclear interaction effects that probably favor the ferromagnetic phase. At lower densities the symmetry of the theory is effectively enhanced to $SU(3)$, and the quantum liquid enters a novel phase, neither ferromagnetic nor nematic.  Our results can serve as a starting point for investigations of the phase dynamics of deuteron liquids, as well as exploration of the stability and dynamics of the rich variety of topological objects that may occur in phases of the deuteron quantum liquid, which range from Alice strings to spin skyrmions to $\mathbb{Z}_2 $ vortices.
\end{abstract}

\maketitle

\section{Introduction}
\label{sec:Intro}
It was recently pointed out \cite{Berezhiani:2010db} that deuterium, when compressed  between atomic and nuclear densities,  becomes a quantum liquid over a large range of temperatures. This comes about through the confluence of a number of factors. First, when the interparticle distance $l$ is smaller than  twice the Bohr radius $a_{0}$, deuterium atoms overlap and are ionized. One then has two separate fluids, one composed of electrons and another composed of deuterons. Due to the Coulomb repulsion between deuterons, which is screened by the electrons only over distances much larger than $l$, the deuterons at zero temperature will crystallize into a lattice. The temperature at which the crystal melts scales as $a_{0} T_{crys}\sim 1/180 \times \alpha (a_{0}/l)$, where $\alpha$ is the electromagnetic coupling constant. On the other hand,  the deuterons Bose-condense at  temperatures below $a_{0} T_{cond}\sim 4\pi^{2}/3 \times (M a_{0})^{-1} (a_{0}/ l)^{2}$, where $M$ is the deuteron mass~\cite{Gabadadze:2009dz}. (We work with units where $k_{B} = \hbar=c = 1$.) Therefore, at high enough densities ($l\alt a_0$), there is a range of temperatures, $T_{crys} < T < T_{cond}$, where a quantum liquid of deuterons should form.  Of course, it must also be the case that the relevant densities are still far from nuclear ones, and Ref.~\cite{Berezhiani:2010db} pointed out that this is indeed the case for deuterium.  Since deuterons are charged bosons, the quantum liquid will be a superconducting superfluid. This is the regime we address in this paper.

There are two motivations for studying this system.  First, this kind of matter is expected to exist in a layer inside  brown dwarfs that are light enough not to ignite deuteron fusion\cite{Berezhiani:2010db}. Second, it may be created in terrestrial laboratories through shock compression \cite{PhysRevLett.78.483, PhysRevLett.99.185001}, experiments using inertial confinement or other techniques \cite{Badiei200970,Andersson20093067}.   To get an estimate of the pressures required, suppose that the pressure is dominated by the electron degeneracy pressure.  This should be reasonable once the deuterium is ionized, which one would expect to take place once $l \sim 2a_{0}$.  The pressure $P$ is then given by
$
P \approx 1.8\times10^{3}\left(\frac{2a_{0}}{l}\right)^{5} \mathrm{GPa}
$,
where we define the interparticle distance as $l \equiv n^{1/3}$, and $n$ is the particle density.  Since $T_{cond}$ exceeds $T_{cryst}$ once $2a_{0}/l \sim 1$, deuterium should become a quantum liquid of deuterons once the pressure reaches  $P_{0} \gtrsim 1.8\times 10^{3}\, \mathrm{GPa}$.  $P_{0}$ sensitively depends on the estimates for the ionization density and $T_{cond}, T_{cryst}$, so that $P_{0}$ should be viewed as only a rough estimate.  For instance, recent quantum molecular dynamics simulations suggest that deuterium may already be ionized at densities of $\sim 400 \,\mathrm{GPa}$~\cite{Wang:2010fk}, an order of magnitude less than naively setting $l =2 a_{0}$ above would suggest.  

Current diamond anvil cells reach a pressure of a few hundred GPa, as do inertial confinement experiments\cite{RevModPhys.66.671,1998Sci...281.1178C,2000PhRvL..84.5564C,2005PhRvB..71i2104B,2009PhRvB..79a4112H}.  Both are a factor of $10-100$ shy from our estimate of the pressures necessary for the appearance of the phases which we will be discussing.  However due to the importance of the subject for energy production and other applications, one can hope that this gap will be closed in the future.  Our goal is to explore some of the unusual properties of this kind of matter as a guide to these experiments.

Since we are interested in phenomena occurring at length scales much larger than the size of the deuterons, the system can be described by a Lagrangian with point-like deuterons and electrons interacting via electromagnetism.  In the regime of interest, the momenta of the deuterons are very small compared to their mass $M \sim 2\, \mathrm{GeV}$, so the deuterons are non-relativistic.  Since the deuterons are charged spin-1 bosons, they can be described by a complex 3-vector field $\mathbf{d}$.  The non-relativistic Lagrangian governing this system is
\begin{align}\label{eq:L}
\mathcal{L}+\mu\mathcal{N}  &=
\ded \Big(iD_0+\mu+\frac{\vec{D}^2}{2M} \Big)\de + \bar\psi (i\dslash +m+\mu_e\gamma^0)\psi \nn \\
&-\frac{1}{4}F_{\mu \nu}F^{\mu \nu}+\frac{e g}{2M}\mathbf{B} \cdot (\ded\times\de)+\cdots,
\end{align}
where $D_{\mu} = \partial_{\mu}+ie A_{\mu}$, and $\psi$ and $A_\mu$ are, respectively, the electron and photon fields, $\mathbf{B}$ is the magnetic field, $g\approx 0.857$ the magnetic moment of the deuteron in Bohr magnetons, $m$ is the electron mass, and $\mu, \mu_e$ are the chemical potentials of the deuterons and electrons; $\ded \times \de = \epsilon^{i j k} d_{j}^{*} d_{k}$ in a slight abuse of notation.  The Lagrangian has an $O(3)$ global rotation symmetry, and a $U(1)$ electromagnetic gauge symmetry\footnote{The system of course also has a  global deuteron-number symmetry, but this symmetry is included in $U(1)_{EM}$.}.   Not shown explicitly in the Lagrangian are the deuteron interaction terms, and terms with more derivatives, the form of which is constrained by the requirement of Galilean invariance.  The inclusion of interactions among deuterons is complicated, since two deuterons at rest lie above the threshold for ${}^{3}He+n, {}^{3}H+p$ production, so that all of these different nuclei need to be included in an effective field theory. This can be done following the approach  to such effective theories described, for instance, in Refs.~\cite{Bedaque:2003wa, Bedaque:2002mn}. These  contributions are subleading but will, in fact,  play a decisive role below. We postpone a further discussion of the deuteron-deuteron interactions until Sec.~\ref{sec:Symmetries}. 

\section{Phases and the effective potential}
\label{sec:EffPot}

At low temperatures ($T<T_{cond}\approx 10^6 K$ for $l\approx 10 a_0$), the deuterons condense, leading to a ground state expectation value for the $\de$ field.  We assume that the condensate is spatially homogeneous, and leave the analysis of non-homogenous phases for a later publication. The  expectation value of $\de$ can be split into real and  imaginary parts as $\langle \de \rangle = \ab+i \bb$, so that $\ab$ and $\bb$ are real 3-vectors. Depending on the relative orientation of  $\ab$ and $\bb$, one finds two different phases, sometimes named  `ferromagnetic' and `nematic' phases in the condensed matter/cold atoms literature. If the condensate is such that $\ab$ and $\bb$ are parallel, we have the nematic phase
\beq
 \langle \de \rangle = \ab\, e^{i\alpha}.
\eeq 
In the nematic phase, a rotation of $\ab$ by $\pi$ can be undone by sending $\alpha \rightarrow \alpha +\pi$, so the order parameter is a  `director' rather than a vector.  If instead  $\ab$ and $\bb$ point in different directions, we have the ferromagnetic phase
\beq
 \langle \de \rangle = \ab+i \bb,  \;\ab\times\bb\neq 0,
\eeq 
where the spin $S = \frac{-i}{2} \ded \times \de =  \ab \times \bb$ of the condensate is non-vanishing.

Our first task will be to decide which of these two phases is actually realized in the deuteron liquid.  To do this, one must evaluate the effective potential $V(\ded,\de)$ and minimize it.  ($V(\ded,\de)$ should be seen as a function of the expectation values  $\langle \de \rangle $ and $\langle \ded \rangle$;  to simplify notation we will drop the brackets except where this could cause confusion, but it is important to keep in mind that the arguments of $V(\ded,\de)$ are classical, not quantum, fields.)  

The effective potential $V(\ded,\de)$ is given by the sum of all one-deuteron irreducible diagrams with $\de, \ded$ external legs and zero external momentum\cite{Peskin:1995ev}. Due to rotation and $U(1)_{EM}$ gauge symmetry it can only depend on the combinations $(\ded \cdot \de)=a^2+b^2$ and $(\ded \cdot \ded) (\de \cdot \de)=(a^2-b^2)^2+4(\ab \cdot \bb)^2$.  If we retain only the terms  up to quartic order in the effective potential we have:
 \beq
 \label{eq:EffPotExample}
 V(\ded,\de) = -\mu \ded\de + c_1 (\ded\de)^2 + c_2 \ded^2 \de^2.
 \eeq 
For $c_2>0$, $V(\ded,\de)$ is minimized by $a=b, \ab \cdot \bb=0$, that is, we have the ferromagnetic phase. In the opposite case,  $c_2<0$,   $V(\ded,\de)$ is minimized by $\ab$ parallel to $\bb$, which yields the nematic phase. Taking into account terms of higher order in $\ded,\de$ other intermediate angles between $\ab$ and $\bb$ may be favored.

\begin{figure}[t]
  \centering
  \includegraphics[width=8cm]{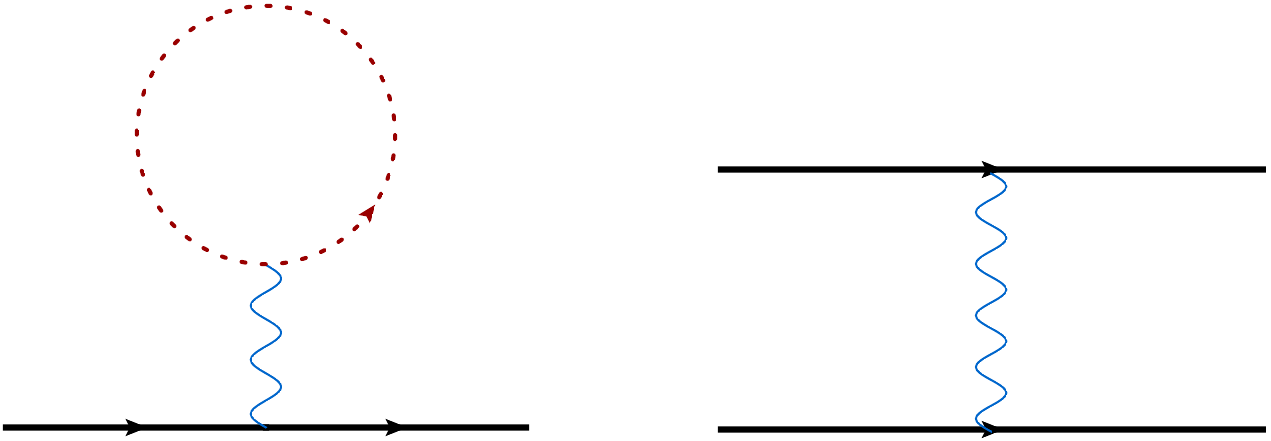}
\noindent
\caption{Example of two graphs  whose derivatives evaluated at $\ded\de=n$ cancel in the effective potential. The dotted lines are electrons, the solid lines deuterons and the wavy line an electron-dressed Coulomb photon line.}
\label{fig:dangling}
\end{figure}  
 
We will now argue that, at leading order in the expansion parameter $l/a_0$ ($a_0=1/\alpha m_e$ is the Bohr radius), the effective potential is given by an infinite series of one-loop diagrams containing only screened Coulomb photons.  Since the dominant Coulomb interaction is spin-independent, this amounts to the claim that $c_{1}\gg c_{2}$ in Eq.~(\ref{eq:EffPotExample}). To show this we have to argue that all other diagrams are suppressed by powers of $l/a_0$ or by the deuteron mass $M$.  What complicates this argument is that the estimate of any diagram we neglect depends on the value of the chemical potential $\mu$. However, recall that $\mu$ serves to enforce charge neutrality, since it is is the chemical potential necessary for the existence of the same density of deuterons and electrons.  This means that one can only compute $\mu$ \emph{after} $V(\ded,\de)$  is known. We break this impasse by positing a certain estimate for $\mu$, and then making sure it is self-consistent.  Specifically, we suppose that $\mu \sim \alpha/l$, and compute $V(\ded,\de)$ assuming this estimate of $\mu$.  We then use the calculated $V(\ded,\de)$ to compute $\mu$ by demanding that charge neutrality be enforced, which is the condition that
 \beq
 \label{eq:neutrality}
\left. \frac{\partial V(\ded,\de)}{\partial(\ded\de)}\right|_{\ded\de=n} =0,
 \eeq 
 where $n =  1/l^3$ is the electron density, and check for self-consistency. 
 
Our first step is to integrate out the Coulomb photons  $A_0$ (ignoring the magnetic photons for the moment). The result is
 \begin{align}
 \label{eq:NoCoulombPhotons}
 &\mathcal{L}+\mu\mathcal{N}  =
\ded\Big(i\partial_0+\mu+\frac{\vec{\nabla}^2}{2M}\Big)\de \\
&+\int d^3r' (\ded\de(r)-\bar\psi\gamma^0\psi(r))  V_{c}(r-r') (\ded\de(r')-\bar\psi\gamma^0\psi(r'))\nn\\
&+\cdots \nn,
 \end{align}
 where $V_{c}$ is the Coulomb potential screened by the presence of the electron Fermi sea;  the Fourier transform of $V_{c}(q)$ at small momenta ($q\ll 1/l$) is
 \beq
 V_{c}(q) = \frac{\alpha}{q^2+m_s^2}
 \eeq
 with the Debye screening mass $m_s^2\sim\alpha m_e/l = 1/a_0 l$ \cite{Fetter:1971fk}.   We do not include magnetic interactions at this stage since they are suppressed by powers of $1/M$, and thus one would expect their contribution to the effective potential to be suppressed.  Of course, this needs to be checked self-consistently, and the effects of magnetic interactions are discussed in Sec.~\ref{sec:Symmetries}.   
 
 Eq. (\ref{eq:neutrality}) implies a cancellation between certain diagrams.   Consider the two diagrams in Fig.~\ref{fig:dangling}. The first one contributes to the $\ded\de$ term in $V(\ded,\de)$, while the second contributes to the $(\ded\de)^2$ term. Their derivatives evaluated at $\ded\de=n$, however, cancel against each other:
 \beq
 \frac{\partial}{\partial(\ded\de)} \left( - \frac{\alpha n}{m_s^2}\ded\de + \frac{\alpha}{2m_s^2}( \ded\de)^2\right)=0.
 \eeq
The same cancellation occurs for any graph containing a ``dangling" deuteron line connected to the rest of the diagram by only one photon line, and thus we can disregard them from now on. 

\begin{figure}[!t]
  \centering
  \includegraphics[width=8cm]{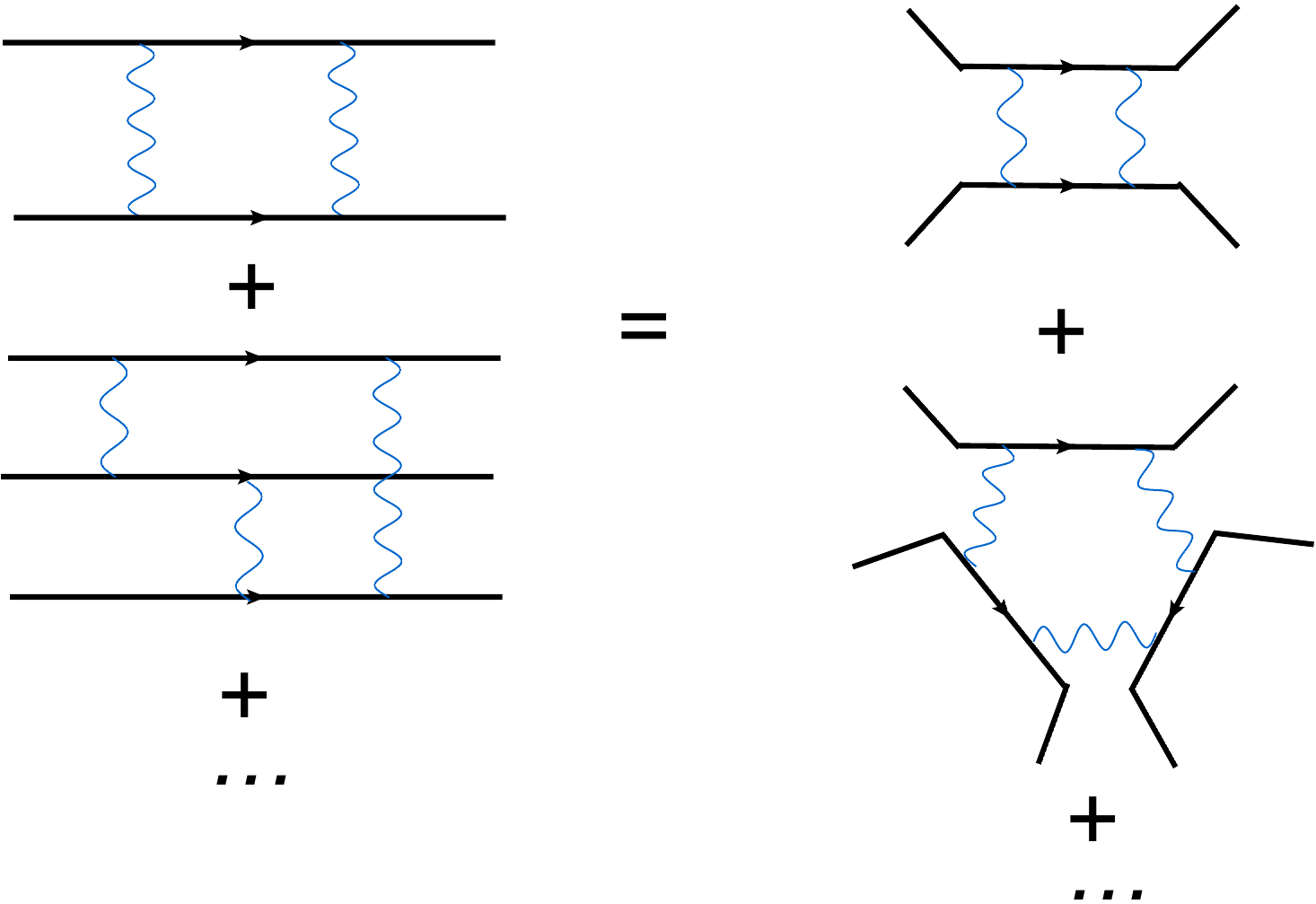}
\noindent
\caption{Class of graphs contributing to the leading order effective potential displayed in two different ways.}
\label{fig:oneloop}
\end{figure}

This cancellation has a simple physical interpretation. Each deuteron interacts with the average charge of the background of other particles. The charge of this background, however, vanishes as the electron charge cancels that of the deuterons. Any shift in the energy will come from charge density {\it fluctuations}, and these are described by two (or more) photon exchanges. An example of a contribution of this kind is shown in the first graph of  Fig.~\ref{fig:oneloop}. More complicated contributions involving more deuteron and photon lines, however, contribute equally and need to be resummed.  
 
 Fortunately, the resummation involved is fairly standard, and amounts to the computation of the one-loop effective potential~\cite{Peskin:1995ev}. Physically, it can be seen as the sum of zero-point energies of the deuteron quasiparticles, including the Coulomb interaction with the condensate.  Alternatively, it can be seen as the sum of the zero-point energies of the photon quasi-particles dressed by interactions with the electrons and the deuteron condensate. The one-loop effective potential $V_{1}$ can be obtained by writing the deuteron field as $\de =  \langle \de \rangle+\chi$, keeping only the terms quadratic in the fluctuations $\chi^\dagger, \chi$ and performing the resulting functional Gaussian integral. The result has the well known ``trace log" form that gives
 \beq\label{eq:oneloopintegral}
 \left. \frac{\partial V_1(\ded,\de)}{\partial(\ded\de)} \right|_{\ded\de=n}= -\mu 
 \;+ \frac{2}{3}  \int \frac{d^3q}{(2\pi)^3} \frac{V_c(q) (n V_c(q) +2 \epsilon_q)}{E_q},
 \eeq  
where $\epsilon_{q}=q^2/2M - \mu$, $E_q^2 = \epsilon_q^2 + (4/3) n V_c(q) \epsilon_q$.  The electron density $n=1/l^3$ appears here because, up to terms suppressed by higher powers of $\alpha$, we can make the substitution $\bar\psi\gamma^0\psi\rightarrow n$ in Eq.~(\ref{eq:NoCoulombPhotons}).   Note that the expansion of $V_1$  in powers of $V_c(q)$ gives the sequence of diagrams portrayed in Fig.~\ref{fig:oneloop}. 

The integral appearing in Eq.~(\ref{eq:oneloopintegral}) would be infrared divergent if not for the electron screening (the $m_s^2$ term) and condensate screening, which appears as the term proportional to $n V_{c}(q)$ in the denominator. That means that the integral will be dominated by small values of $q$, on the order of $q\sim m_s$. When $l \ll a_0$ and  $\mu \sim \alpha/l$ we have that $\epsilon_q \sim \mu \sim \alpha/l \sim ( l/a_0 ) n \alpha/m_s^2 \ll n V_{c} $, and so $\epsilon_{q}$ can be neglected in the numerator. Similarly, in the denominator, we use the hierarchy of scales 
\beq
q^2/M  \sim m_s^2/M \ll \alpha/l \sim \mu \ll n \alpha/ m_s^2 \sim n V_{c}. 
\eeq
This means that the condensate-screening $nV(q)$ term in $E_{q}$ in the denominator gives the dominant contribution in the regime of interest.  We see then that the infrared divergences in the potential are partially cutoff by  the Debye screening mass $m_s$, and partially by the effect of the condensate. 

Using the results above, we can now estimate the integral in Eq.~(\ref{eq:oneloopintegral}) and find the value of the chemical potential necessary to keep the deuteron density equal to the electron density.  This condition turns out to imply that
 \beq
 \mu \sim \frac{n \alpha^2}{\sqrt{\mu}} \sqrt{\frac{m_s^2}{n\alpha}} \frac{1}{m_s} \Rightarrow \mu \sim \frac{\alpha}{l},
 \eeq
so that our estimate of $\mu$ is self-consistent, as previously advertised. 

Notice that each individual diagram that was summed above is parametrically larger than their sum. In fact, the one-loop diagram with $n$ photon lines is of order $(\alpha/l^4)(a_0/l)^{n-3/2}$ so they are larger the more photon lines they have. Their sum, however, is smaller, of order $\alpha/l^4$. This is because the infrared divergences are actually cut off by the scale set by the deuteron condensate, an effect not included at any finite order in perturbation theory, instead of the scale $\mu$ appearing in individual Feynman diagrams.  
 
\begin{figure}[!t]
  \centering
  \includegraphics[width=5cm]{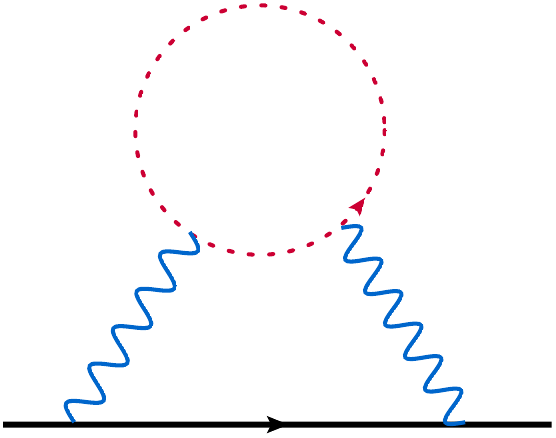}
\noindent
\caption{An example of a sub-leading graph. The dotted line represents an electron.}
\label{fig:electronblob}
\end{figure}

What remains to be done now is to argue that other graphs would, in the regime considered, give contributions smaller than the ones that we kept.  As mentioned before, these estimates hinge on the value of $\mu\sim \alpha/l$ and finding this value was indeed  the main motivation for the effective potential argument above.
The suppression of the remaining diagrams is best argued through some examples. Consider, for instance, adding one photon line to the one-loop diagrams of Fig.~\ref{fig:oneloop}. The effect is to substitute one $V_{c}(q)$ by a two-photon ladder with zero incoming (relative) momentum and $q\sim m_s$ outgoing momentum. This ladder diagram can be estimated as
 \beq
 \int d^3k \frac{\alpha}{q^2+m_s^2}\frac{\alpha}{(q-k)^2+m_s^2}\frac{1}{\epsilon_k}
 \sim
 \frac{\alpha^2}{\mu m_s} 
 \sim
 \frac{\alpha}{m_s^2}\sqrt{\frac{l}{a_0}}
 \eeq 
 and, consequently, gives a smaller contribution compared to the leading diagrams by a factor of $\sqrt{l/a_0}$. Graphs including electrons are also suppressed. For instance, let us look at at graph in Fig.~\ref{fig:electronblob}. Its contribution to the effective potential can be estimated as
 \beq
 \int d^3q \left(  \frac{\alpha}{q^2+m_s^2}\right)^2 \frac{m}{l} \sim \frac{\alpha}{l^4}\sqrt{\frac{l}{a_0}},
 \eeq
 where we used the fact that the electron loop, at small momenta $q\sim m_s$ is of order $\sim m/l$~\cite{Fetter:1971fk}. This contribution is again suppressed by  $\sqrt{l/a_0}$ compared to the leading order. The picture emerging then is very similar to the high density limit of the jellium model, which is a charged Bose gas with a fixed, non-dynamical background of negative charges~\cite{Foldy:1961kx,Brueckner:1967vn}, modified by the electron screening of the Coulomb force. 

\section{Symmetries of the deuteron liquid}
\label{sec:Symmetries}

The most important consequence of identifying the one-loop Coulomb diagrams as giving the leading contribution to the effective potential is that the effective potential is a function of $\ded\de$ only, not of $\ded^2 \de^2$. This  can be seen from the fact that all the incoming deuteron lines are contracted with outgoing deuteron lines on the graphs in Fig.~\ref{fig:oneloop}. Physically, this follows from the spin-independence of the Coulomb forces. This implies that the symmetry group of our theory, which is $O(3)\times U(1)_{EM}$, is effectively enlarged to $SU(3)\times U(1)_{EM}$~ 
\footnote{By writting $\de=\mathbf{a}+i \mathbf{b}$ and considering $\mathbf{a}$ and $\mathbf{b}$ as the components of a real 6-dimensional vector, one may think  that the symmetry group becomes $O(6)$. This is indeed true for terms of the form $(\ded\de)^{n}$, but it is \emph{not} true for the kinetic term. 
$O(6)$ would be a symmetry of the kinetic term for a {\it relativistic} bosonic theory.  A familiar example of this occurs in the Higgs sector of the Standard Model, which has the symmetry $SU(2) \times SU(2) \simeq SO(4)$, where the first of the $SU(2)$'s is gauged and the second is the custodial $SU(2)$ symmetry.  However, a non-Lorentz invariant kinetic term breaks the O(2N) symmetry to SU(N), so that the symmetry of the deutron liquid is $SU(3)\times U(1)_{EM}$ and not $O(6) \times U(1)_{EM}$.}.

\begin{figure}[t]
  \centering
  \includegraphics[width=5cm]{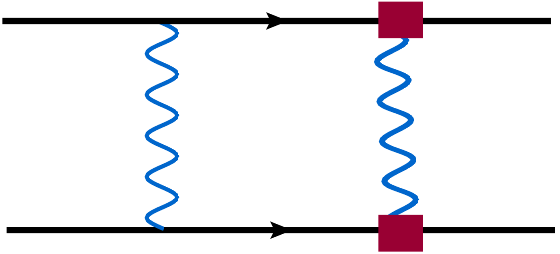}
\noindent
\caption{Leading electromagnetic $SU(3)$ breaking effect. The square vertex represents a magnetic dipole interaction.}
\label{fig:su3break}
\end{figure}  

We now examine the size of the $SU(3)$ breaking effects. The leading electromagnetic $SU(3)$ breaking diagram is shown in Fig.~\ref{fig:su3break}. This diagram can be estimated as
\begin{align}
V^{EM}_{break} &\sim \int d^3q \left\{ \frac{\alpha}{q^2+m_s^2} \frac{\alpha}{M^2} \right.\nn\\ 
&\times \left. \frac{q^iq^j}{q^2+\frac{\pi^{2}}{2}m_{s}^{2}\frac{q_{0}}{q}} (d^{*}_{i} d_{j}\de^2 + d_{i} d^{*}_{j} \ded^2 -2 d^{*}_{i} d_{j} \ded \cdot \de) \right\}\nn\\
\end{align}
The loop integral above, which would be IR-divergent if not for static and dynamic screening effects, contains three scales: $\mu, m_{s}$, and the scale associated with the Landau damping  of the magnetic photon, $q_{m} \sim (m_{s}^{2}\mu)^{1/3}$. For $q \gg q_{m}$, the propagator of the magnetic photon becomes $q_{i}q_{j}/q^{2} \rightarrow \delta_{ij}$.  For smaller $q$, $q_{m}$ sets the scale for the momenta in the magnetic photon propagator.  Using the fact that $\mu \sim \alpha/l \lesssim m_{s}\sim (a_{0} l)^{-1/2}$,   the estimate of the integral is
\beq
V^{EM}_{break}\sim \frac{1}{M^{2}} q_{m}^{3} \frac{\alpha}{m_{s^{2}}} \frac{1}{\mu} \frac{\alpha}{m_{s}^{2}\mu} q_{m}^{3} \sim \frac{\alpha^{2}}{M^{2}}
\eeq
This contribution would favor the ferromagnetic phase over the nematic phase. Notice that the  integration over $\mathbf{q}$ sets $q^i q^j/q^2 \rightarrow \delta^{ij}/3$, implying that the only part of the dipole potential surviving is, in position space, a $\delta(r)$ function at the origin.   This leads to the surprising conclusion that contact interactions due to nuclear forces may compete with dipole-dipole interactions between the deuterons.   A very similar situation involving dipole-dipole interactions occurs in the physics of $W$ boson condensation in the early Universe \cite{Dolgov:2010gy}. There the $W$ boson contact interactions are fixed by gauge symmetry and are known to favor the ferromagnetic phase. 

In our case the the strong nuclear interaction between two deuterons is more uncertain. In the effective theory we are using, valid at distances larger than the deuteron size ($\approx 5$ fm), the s-wave deuteron-deuteron interaction is described, at lowest order in the momentum expansion by
\beq
\label{eq:su3break_nuclear}
V^{nuc}_{break} \sim \frac{4\pi A_0}{3M}  \ded^2 \de^2 +\frac{4\pi A_2}{3M} \left(  3(\ded.\de)^2 - \ded^2 \de^2\right).
\eeq 
The term proportional to $A_0$ ($A_2$) contributes to the spin channel $S=0$ ($S=2$) deuteron-deuteron channel. In fact, in our effective theory, the s-wave scattering lengths in the spin $S=0,2$ channels  would be given by the terms in Eq.~(\ref{eq:su3break_nuclear}) dressed by Coulomb photons with momenta within the range of validity of the theory ($Q\alt 1/5\, \mathrm{fm}^{-1}$), which are explicitly included in Eq.~(\ref{eq:L}). Thus, the coefficients $A_0, A_2$ above are the deuteron-deuteron scattering lengths in the \emph{absence} of soft Coulomb interactions, but including the effect of hard photons ($Q\agt 1/5\,\mathrm{fm}^{-1}$).

What complicates the description of deuteron-deuteron scattering is the existence of channels below the deuteron-deuteron threshold, namely, $n+{}^3He$ and $p+{}^3H$. One way of dealing with these additional channels is to include  $n,p,{}^3He$ and ${}^3H$ fields explicitly in the effective theory, together with their respective couplings, and perform a coupled-channels calculation in the medium. Most likely this is unnecessary. The presence of these open channels will make the deuteron-deuteron phase shifts complex, but this can be rigorously taken into account by taking the parameters $A_0, A_2$ to be complex. This technique is used, for instance, in non-relativistic QCD where a quark-antiquark system can annihilate into energetic gluons \cite{Bodwin:1994jh}. 

In any case, one expects $A_{0}$ and $A_{2}$ to have very small imaginary parts. To see this, note that the deuteron-deuteron initial state can have $S=0$ or $S=2$.  First, consider the $S=2$ deuteron-deuteron initial state. The intermediate state can have only $S=0$ or $S=1$, so it will have some non-vanishing angular momentum. At small energies, mixing with higher partial waves is suppressed, so the scattering length in the $S=2$ channel should have only a small imaginary part.  Next, consider the $S=0$ initial state.  This initial state can mix with the intermediate $S=L=0$ state. However, the wave function configuration of the initial and intermediate states is very different, so this mixing should be very small.  

In fact, a model calculation of the $S=0$ deuteron-deuteron scattering length gives $A_{0} = 4.91\pm0.02 +i( -0.0115 \pm 0.0001)$ fm~\cite{PhysRevC.58.58}. In our effective theory, the values of $A_0$ and $A_2$ are the scattering lengths in the spin $S=0,2$ channels in the absence of Coulomb interactions. More precisely, $A_{0}$ and $A_2$ subsume electromagnetic interactions at distances smaller than $\approx 5$ fm (or photons with momenta $\gg 1/5$ fm); the contribution from softer photons is included explicitly in Eq.~(\ref{eq:L}). 

 If the deuteron forces were spin-independent, $A_0$ would equal $A_2$, and we would have an effective potential with the enhanced $SU(3)$ symmetry. 
Current few-body nuclear technology is available for a realistic calculation of these parameters, but we are aware of only two published works. 
One model calculation, using a simple Malfliet-Tjon potential \cite{Malfliet:1970vd} gives $A_0=10.2 $ fm and $A_2=7.5$ fm~\cite{Filikhin:2000uq}.These values for the scattering lengths lead to a ferromagnetic phase.  Ref.~\cite{PhysRevC.58.58} computes only $A_0$ giving the value quoted above ($A_{0} = 4.91\pm0.02 +i( -0.0115 \pm 0.0001)$ fm). 

A. Deltuva was kind enough to use the methodology described in \cite{PhysRevC.75.014005}, \cite{PhysRevC.76.021001} to compute the required scattering lengths {\it without electromagnetic forces} at our request~\cite{DeltuvaPrivate}. The results he finds are $A_0=5.35, 5.13, 4.87$ fm and $A_2=3.16, 3.16, 3.18$ fm,  respectively for the AV18, CDBonn and INOY04  potentials. The imaginary parts are all, as expected, of the order of $1\%$. None of these calculations include the effect of three-body forces, known to contribute a small (about 5\%) to the binding energy of small nuclei or the hard photon exchange. There is a high degree of universality in low-energy few-nucleon reactions -- understood from the point of view of effective field theories ---  so the discrepancy of this result with the one in \cite{Filikhin:2000uq} can hardly be blamed on the difference between models used.  In any case, an estimate of the size of these scattering lengths can be made by just assuming a scattering length comparable to the ``size" of the deuteron itself, namely, a few Fermi.  This means that $V^{nuc}_{break}\sim A/M$ with $A \approx 5$ fm is actually larger than $V^{EM}_{break}$, since
\beq
\frac{\alpha^{2}}{M^{2}} \ll \frac{A}{M}
\eeq
given that $1/M \sim 10^{-1} fm$. 

\begin{figure}[t]
  \centering
  \includegraphics[width=4cm]{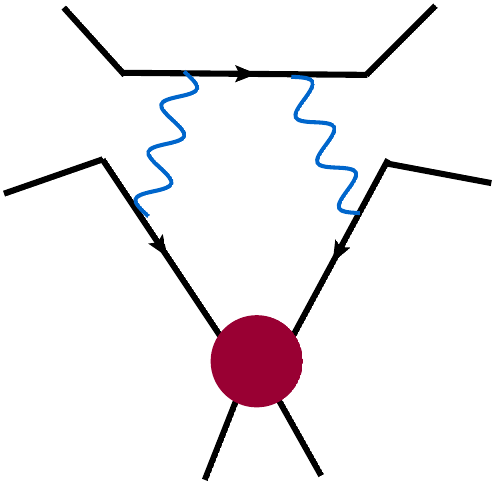}
\vskip 0.15in
\noindent
\caption{Leading nuclear $SU(3)$-breaking contribution to the  effective potential. The contact term is a spin-dependent nuclear interaction.}
\label{fig:nuclear_perturbation}
\end{figure}  

It is natural to ask whether the nuclear interactions can favor angles between $\ab$ and $\bb$ that are \emph{not} $0$ or $\pi/2$.  As noted after Eq.~(\ref{eq:EffPotExample}), such an effect would have to come from terms with more than four powers of the $\de$ fields in the effective potential.  Of course, the one-loop effective potential contains terms with all powers of $\ded \cdot \de$, and it is spin-independent. The leading spin-dependent contribution to the effective potential can be obtained by replacing one of the Coulomb photons in the diagrams leading to the one-loop effective potential with a nuclear contact interaction between the deuterons, as shown in Fig.~\ref{fig:nuclear_perturbation}.  However, it is not hard to see that this merely gives a contribution to the potential of the form
\beq
\frac{4\pi A_0}{3M}  f(\ded \cdot \de) \ded^2 \de^2 +\frac{4\pi A_2}{3M} g(\ded \cdot \de) \left(  3(\ded \cdot \de)^2 - \ded^2 \de^2\right),
\eeq
for some functions $f$ and $g$ that depend only on $\ded \de$.  Thus at this order the effective potential would pick out a phase with either $\ab \times \bb = 0$ or $\ab \cdot \bb = 0$.  To get other angles between $\ab$ and $\bb$, we must have at least \emph{two} nuclear contact interactions in the effective potential.  This sort of contribution is highly suppressed by powers of $1/M$.   So to leading order, the nuclear interactions drive the deuteron liquid into either the nematic phase, or the ferromagnetic phase with $\ab \cdot \bb = 0$.  

Finally, one may wonder how it is that nuclear effects can come to affect the deuteron liquid, since the strong interactions are short-ranged, and naively the deuterons are prevented from getting close to each other by Coulomb repulsion, which is only screened on distances larger than $a_{0}$.  However, in the regime of interest the dynamics of the deuterons cannot be thought of in terms of collisions between classical particles.  The deuterons are in a quantum condensate, so that their wave-functions overlap, and they are highly off-shell, making classical intuition about scattering difficult to apply.  One might have thought that electromagnetic dipole interactions would be long-ranged compared to the nuclear interactions, but as it turns out they yield only contact interactions to leading order, due to angle-averaging and the spatial homogeneity of the condensate.  Since the strength of the electromagnetic SU(3)-breaking contact interactions turns out to be much smaller than that of the nuclear ones, it turns out that nuclear interactions can play an important role in the dynamics of the deuteron liquid.

\section{Finite temperature effects}
\label{sec:FiniteT}

Since the size of the $SU(3)$ breaking effects is suppressed by powers of $1/M$, one may wonder whether they will actually have any physical consequence after the small, but finite, temperature effects are included. One way of estimating this is to compare the energy density required to move the $\mathbf{a}$ and $\mathbf{b}$ fields from parallel to orthogonal directions
\beq
\mathcal{E}_\theta \sim \frac{4\pi A}{3M}\frac{1}{l^6} \cos^2\theta
\eeq 
to the thermal energy density available in the system
\beq
\mathcal{E}_T \sim \frac{T}{l^3}.
\eeq
When $\mathcal{E}_T$ is greater than $\mathcal{E}_{\theta}$, one would expect that SU(3)-breaking effects get washed out.  We then find that 
\beq
\frac{\mathcal{E}_\theta}{\mathcal{E}_T} \sim   \frac{4\pi A}{M a_0^3 T}  \left( \frac{a_0}{l}\right)^3
\sim
 \frac{10^{-6}}{a_{0}T}  \left( \frac{a_0}{l}\right)^3.
\eeq 
At the typical temperatures (of order $10^{5} K \Rightarrow a_{0}T \sim 1$) that we are interested in, where $T$ is larger than the crystallization temperature but smaller than the condensation temperature, the criterion above splits the interesting range of densities into two regions.
	
At larger densities ($l \lesssim 10^{-2}a_0$) the $SU(3)$-breaking effects are important and determine the phase of the system. With $A_0 > A_2$ the ferromagnetic phase is favored. Once the global $SU(3)$ symmetry is broken, the symmetry of the system prior to deuteron condensation is $O(3)\times U(1)_{EM}$.  In the presence of a ferromagnetic condensate $\langle \de\rangle = \ab +i \bb$, which has $\ab \cdot \bb=0$ and $a = b$, the symmetry breaks to to a global $U_{z-Q}(1)$ subgroup combining a rotation around the $\ab \times \bb$ `z' axis and an opposite phase rotation. The vacuum manifold is then $(O(3)\times U(1)_{EM})/U_{z-Q}(1) \simeq O(3)$. This pattern of symmetry breaking also occurs in cold atom optical traps of spin-1 atoms\cite{PhysRevLett.81.742}.  

If it turns out that $A_{2} > A_{0}$, the nematic phase would be favored instead.  In this phase the $O(3)\times U(1)_{EM}$ symmetry is broken to $O(2) \ltimes Z_{2}$.  If the condensate points along the $z$ direction, the $O(2)$ consists of rotations around the z-axis, as well 2D-parity transformations $(x,y) \rightarrow (-x,y)$, while the $\mathbb{Z}_{2}$ is realized as a rotation of the condensate by $\pi$ around the $x$ or $y$ axes, followed multiplication by $e^{i\pi} \in U(1)_{EM}$.  Thus the vacuum manifold is $(O(3) \times U(1)_{EM})/(O(2) \ltimes \mathbb{Z}_{2} )$. 

At lower densities, which are more accessible to experiments and observation, $\mathcal{E}_\theta/\mathcal{E}_T \ll 1$ and the $SU(3)$-breaking effects are relatively unimportant. In that case the symmetry of the theory is effectively enhanced from  $SO(3)\times U(1)_{EM}$ to  $SU(3)\times U(1)_{EM}$.   Using $SU(3)\times U(1)$ transformations it is possible to go from the ferromagnetic phase with $\ab \cdot \bb = 0$ to the nematic $\mathbf{a}\times\mathbf{b}=0$ phase.  In this regime, the deuteron liquid is in a novel phase, which we will refer to as the $SU(3)$ phase; it is neither a ferromagnetic nor nematic phase.  The vacuum manifold can be found by noting that by a choice of coordinates in the three-complex dimension space we can make $\langle \de \rangle$ to be real and have only its $z$-component non-vanishing.  The symmetry breaking pattern is then $SU(3)\times U(1)_{EM}\rightarrow U(2)$, where $U(2)$ is the subgroup of $U(3)$ that leaves $\langle \de \rangle$ invariant.  The vacuum manifold is then a five-dimensional sphere $(SU(3)\times U(1)_{EM})/U(2)=S^5$.

\section{Topological defects}
\label{sec:Topology}
The vacuum manifolds of theories with charged condensates generally allow for the existence of topological defects.  For the condensation of spin-0 nuclei, this has been examined in Ref.~\cite{Gabadadze:2009qe}, following some foundational work in Refs.~\cite{Gabadadze:2007si,Gabadadze:2008mx,Gabadadze:2008pj,Gabadadze:2009dz,Gabadadze:2009fn,Gabadadze:2009jb,Gabadadze:2009zz}. The vacuum manifolds of the deuteron liquid, in which spin-1 nuclei are condensed, support a rich variety of finite-energy topological defects.  We leave a detailed study of their stability, dynamics, and physical implications to future work, and confine ourselves here to simply sketching how they appear.   

Let us first discuss the situation for the more accessible densities at which $\mathcal{E}_\theta/\mathcal{E}_T \ll 1$, so that $SU(3)$-breaking effects are negligible.  The ground state manifold in this case is simply $(SU(3)\times U(1)_{EM})/U(2)=S^5$, which has trivial homotopy groups $\pi_{k}$ until $k=5$.  Since $\pi_{5}(S^{5}) = \mathbb{Z}$, the effective theory of the deuteron liquid should admit a Wess-Zumino-Witten term with a quantized coefficient \cite{Hill:2009wp}.  It would be interesting to work the physical implications of such a term in this context.

It is sometimes assumed that finite-energy topological defects are classified simply by the homotopy groups of the coset space $G/H$, where $G$ is the full symmetry group of the theory and $H$ is the subgroup of $G$ that leaves the condensate unchanged.  For instance, one might think that if $\pi_{1}(G/H) = 0$ the theory would not support vortex strings.  However, the situation is actually more subtle if only a subgroup $G_{L}$ of $G$ is gauged and breaks to a gauged subgroup $H_{L}$ of $H$.  When this is the case the homotopy groups of the coset space $G_{L}/H_{L}$ of gauge symmetries become important.  When these homotopy groups are non-trivial, various `semilocal' defects are possible\cite{Vachaspati:1991dz, Hindmarsh:1991jq, Hindmarsh:1992yy, Hindmarsh:1992ef,Preskill:1992bf,Achucarro:1999it}.  

In our case, $H_{L}$ is trivial and $G_{L} = U(1)_{EM}$, so the coset space is simply $U(1)$, and the theory supports semilocal vortex strings because $\pi_{1}(U(1)) = \mathbb{Z}$.  To see why a non-trivial $G_{L}/H_{L}$ can result in vortex strings, note that the action of the $U(1)_{EM}$ on $\langle \de \rangle$ defines a circle in the vacuum manifold $S^{5}$.  Motion along this circle costs no energy, and the value of the deuteron field at spatial infinity can wrap the circle in $S^{5}$ defined by the gauge orbit, giving a topologically non-trivial configuration.  The semilocal vortex string cannot unwind even though $\pi_{1}(S^{5}) = 0$, because to do so it would have to leave the gauge orbit at infinity, which would cost an infinite amount of gradient energy.   This discussion makes clear that the non-trivial embedding of the gauge vacuum manifold $G_{L}/H_{L} = U(1) \simeq S^{1}$ in the full vacuum manifold $S^{5}$ gives rise to a fibration of $S^{5}$ over $\mathbb{C}P^{2}$ with fiber $S^{1}$.   Since $\pi_{2}(\mathbb{C}P^{2}) = \mathbb{Z}$, the $SU(3)$ phase of the deuteron liquid can contain global monopoles;  these monopoles live on the ends of semilocal vortex strings\cite{Hindmarsh:1992yy,Hindmarsh:1992ef}.  Semilocal defects are not always stable, and we leave an investigation of their stability and physical effects using the power counting developed in this paper  to later work.  

Let us now consider the two possible phases of the deuteron liquid at higher densities, $l \lesssim 10^{-2}a_0$), when the $SU(3)$-breaking effects are important and determine the phase of the system.  First, consider the ferromagnetic phase.  The symmetry  $O(3)\times U(1)_{EM}$ is broken down to a global U(1) symmetry;  as noted previously the global U(1) is generated by a linear combination of generators of the global $O(3)$ symmetry and the local $U(1)_{EM}$ symmetry.  Before discussing semilocal defects, consider the usual kind of topological defects,  which can be classified by the homotopy groups of the full vacuum manifold.  The vacuum manifold in the ferromagnetic phase is just $O(3)$.  Since $\pi_{1}(O(3)) = Z_{2}$, the ferromagnetic phase supports topologically stable vortices, but ones that are rather unusual: rather than one vortex for each integer value of the circulation, there is only one kind of topologically non-trivial vortex.  Furthermore, since $\pi_{3}(O(3)) = \mathbb{Z}$, the ferromagnetic phase supports topologically stable `spin Skyrmions'.   (Adapting the usual hedgehog ansatz for the ferromagnetic phase, it is easy to see that the spin direction of the condensate varies in the core of the skyrmion, hence the name.) It is unclear, however, whether the spin Skyrmions is stable against collapse.

From the discussion of the $SU(3)$ phase above, it is not a big jump to realize that the ferromagnetic phase also supports semilocal strings classified by $\pi_{1}(S^{1}) = \mathbb{Z}$.  Furthermore, since the gauge vacuum manifold, which is an $S^{1}$, is again embedded non-trivially in the full vacuum manifold, we find a fibration of $SO(3)$ with fiber $S^{1}$ and base space $O(3)/S^{1}$.  Since  $\pi_{2}(O(3)/S^{1}) = \mathbb{Z}$, the ferromagnetic phase supports global monopoles that sit on the ends of the semilocal strings, as does the $SU(3)$ phase.  

Finally, suppose the SU(3)-broken phase is actually a nematic one.   The symmetry  $O(3) \times U(1)_{EM}$ is broken down to $O(2) \ltimes \mathbb{Z}_{2}$, and the vortices are classified by $\pi_1[O(3) \times U(1)/(O(2)\ltimes \mathbb{Z}_{2})]$, which is not trivial. The vortex strings in this kind of phase are called ``Alice strings'' and are known to have very unusual properties\cite{Schwarz:1982ec,Balachandran:1983pf,Alford:1990ur,Alford:1990mk,2000JETPL..72...46L}. As one travels around the vortex the direction of the condensate changes by $\pi$ (not $2\pi$), since the reversed orientation can be made up by a shift in the phase by $\pi$.  In the relativistic context, particles moving adiabatically around the vortex flip their charge. In our context, they flip their spin.

\begin{figure}[t]
  \centering
  \includegraphics[width=9cm]{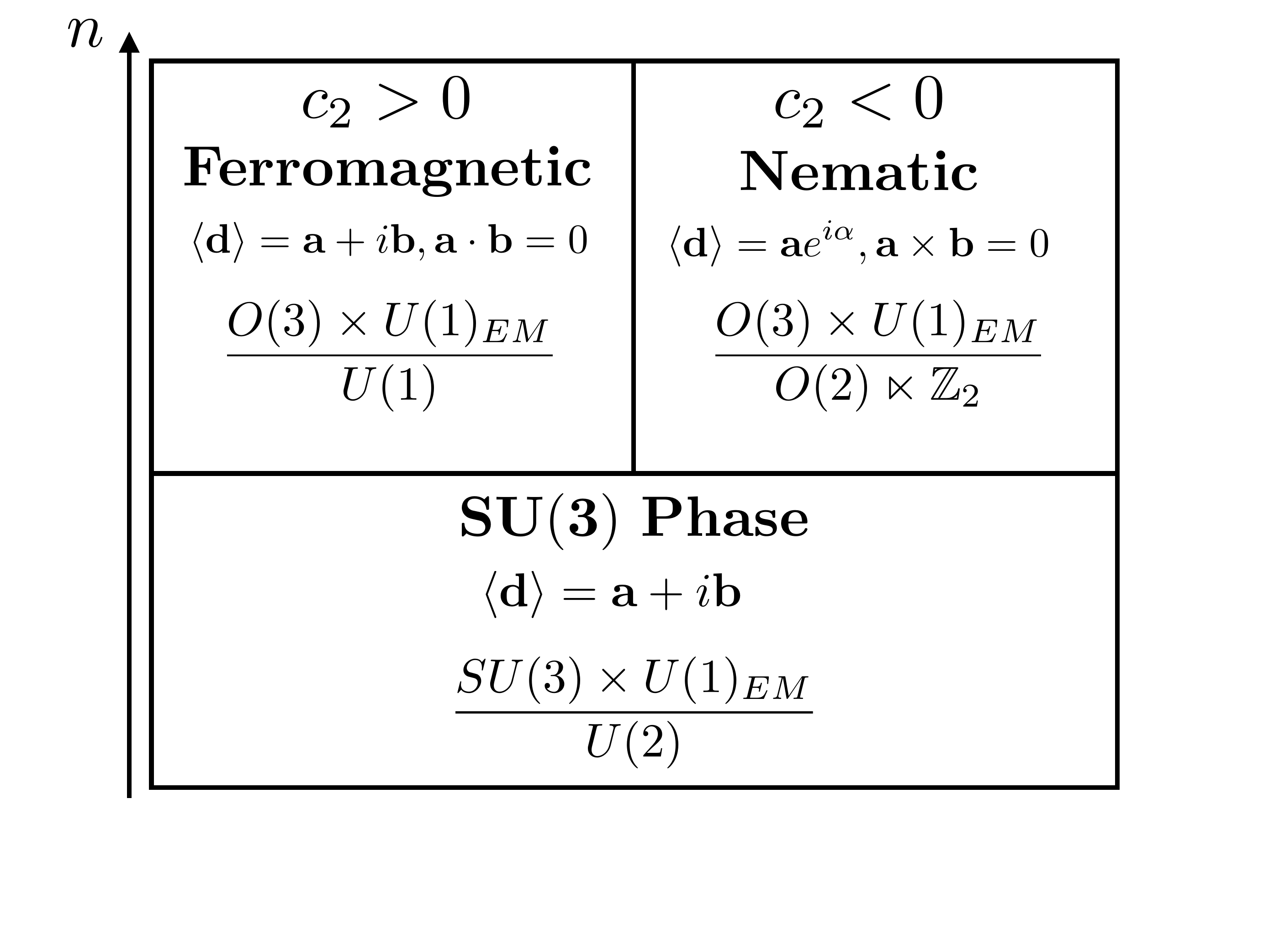}
\noindent
\caption{Table of possible phases, their condensates, and corresponding coset spaces as the density $n\sim 1/l^3$ increases at a fixed temperature $T \sim 10^{5} K$.  The parameter $c_2$ is the coefficient of the  $\ded^2 \de^2$ term in the effective potential, Eq.~\eqref{eq:EffPotExample}, which turns out to be proportional to $(A_0-A_2)/M$.}
\label{fig:phase_table}
\end{figure}  

\section{Conclusion}
\label{sec:Conclusions}

As we have seen above, the deuteron liquid first discussed in Ref.~\cite{Berezhiani:2010db} has a rich phase structure.  To show this, we calculated the one-loop effective potential for the deuteron fields, and showed that to leading order, the theory enjoys an enhanced $SU(3)$ global symmetry.  We then discussed $SU(3)$-breaking effects, and showed that somewhat surprisingly, deuteron-deuteron interactions mediated by the strong force are the dominant source of $SU(3)$-breaking.  However,  as discussed in Sec.~\ref{sec:FiniteT} the importance of the $SU(3)$-breaking effects depends on the density of the system.  At lower densities, the system retains the $SU(3)$ symmetry, while at higher densities the global symmetry of the deuteron liquid is explicitly broken to $O(3)$.  The spontaneous symmetry breaking patterns in these two regimes are very different, as discussed in Secs.~\ref{sec:Symmetries} and  \ref{sec:Topology}.

A brief comment on the relation between the effective field theory used here and the one in Ref.~\cite{Berezhiani:2010db} is probably useful. Our effective theory, valid at distance scales larger than the deuteron radius is `more microscopic' than the one in Ref.~\cite{Berezhiani:2010db}, valid at distances larger than $l$. As such, coefficients that are phenomenological in Ref.~\cite{Berezhiani:2010db} can be computed in our theory. In addition, we considered a more general symmetry breaking pattern (the ferromagnetic phase) than addressed in Ref.~\cite{Berezhiani:2010db}. 

Our results can serve as foundational work for further exploration of these phases.  Let us mention just a few possible future directions.  In our analysis we assumed that the condensate is spatially homogeneous, but of course it is important to check if a spatially-inhomogeneous condensate is also possible.  Next, a more systematic study of finite temperature effects on the deuteron liquid is desirable, as is an explicit computation of the condensation temperature.  It is also essential to better understand the nuclear interactions that break the $SU(3)$ symmetry and drive the deuteron liquid into either the nematic or ferromagnetic phase.

Particularly for applications to astrophysics, it would be useful to understand the magnetic properties of the deuteron liquid.  As part of such an investigation, one would need to explore under the conditions under which the deuteron liquid supports stable vortex strings.  More generally, it will be interesting to investigate the stability and dynamics of the bestiary of topological defects we sketched in Sec.~\ref{sec:Topology}.  

Clearly, deuterium at extreme densities turns out to be an interesting system deserving of further study.

\acknowledgements

We are grateful to E. Berkowitz, T. Cohen, G. Gabadadze, J. M. Laming, L. Platter, J. Rosenberg, J. Sau, and B. Tiburzi for enlightening discussions. We are especially grateful to A. Deltuva for providing us with unpublished results on deuteron-deuteron scattering. We thank the U.S. Department of Energy for support under Grant No. DE-FG02-93ER-40762.

\bibliographystyle{h-physrev}
\bibliography{nuclear_liquids} 

\end{document}